\documentclass[%
 reprint,
 amsmath,amssymb,
 aps,
prl
]{revtex4-1}

\usepackage{graphicx}
\usepackage{dcolumn}
\usepackage{bm}

\begin{document}

\preprint{APS/123-QED}

\title{Simple wave-optical superpositions as prime number sieves}

\author{T. C. Petersen}
\affiliation{%
School of Physics and Astronomy, Monash University, Victoria 3800, Australia.\\
}%
\author{M. Ceko}
\affiliation{%
School of Physics and Astronomy, Monash University, Victoria 3800, Australia.\\
}%
\author{I. D. Svalbe}
\affiliation{%
School of Physics and Astronomy, Monash University, Victoria 3800, Australia.\\
}%
\author{M. J. Morgan}
\affiliation{%
School of Physics and Astronomy, Monash University, Victoria 3800, Australia.\\
}%
\author{A. I. Bishop}
\affiliation{%
School of Physics and Astronomy, Monash University, Victoria 3800, Australia.\\
}%
\author{D. M. Paganin}
\affiliation{%
School of Physics and Astronomy, Monash University, Victoria 3800, Australia.\\
}%
 \email{timothy.petersen@monash.edu.au}

\date{\today}

\begin{abstract}
We encode the sequence of prime numbers into simple superpositions of identical waves, mimicking the archetypal prime number sieve of Eratosthenes. The primes are identified as zeros accompanied by phase singularities in a physically generated wave-field for integer valued momenta. Similarly, primes are encoded in the diffraction pattern from a simple single aperture and in the harmonics of a single vibrating resonator. Further, diffraction physics connections to number theory reveal how to encode all Gaussian primes, twin-primes, and how to construct wave fields with amplitudes equal to the divisor function at integer spatial frequencies. Remarkably, all of these basic diffraction phenomena reveal that the naturally irregular sequence of primes can arise from trivially ordered wave superpositions.    
\end{abstract}

\pacs{PACS numbers: }
\maketitle

We may construe the distribution of primes as a puzzle in physics. Can the seemingly random yet highly orchestrated prime number distribution correspond to states of a physical system \cite{Susan}? Tentative affirmations have arisen from research into connections between physics and number theory, particularly via the Riemann zeta function \cite{RiemannReview}. Building upon Euler's connection between generalized harmonic series and products over primes, Riemann described a Fourier-like analysis to synthesize the prime counting function \cite{Havil}. Hinging upon this construction is the placement of non-trivial zeros of the associated zeta function, which remains elusive \cite{PrimeBook}. 

Diverse studies have revealed the `physics of the Riemann hypothesis', ranging from classical \cite{BilliardsBrunimovich} and quantum billiard balls, quantum scattering and bound states to statistical physics, condensed matter and more \cite{RiemannReview}. Notably, Berry and Keating related the zeros of the Riemann zeta function to eigenvalues in wave systems with classically chaotic trajectories, speculating on the centrality of a simple classical Hamiltonian \cite{BerryKeating}. This approach was recently exploited by studying a non-Hermitian quantization of a Hamiltonian system with real eigenvalues, defined by a postulated maximally broken parity-time symmetry, to imply validity of the Riemann hypothesis \cite{Bender}.   

Diffraction physics is also rich in number theory connections such as Cantor set fractals arising from solitons in non-linear optical fibers \cite{CantorSet}. The discovery of complex exponential Gauss sums in the fractional Talbot effect, which arise in analytic number theory, is particularly pertinent \cite{BerryKlein, Talbot}. Integer factorization was recently achieved in a wave optic experiment exploiting the Talbot effect \cite{TalbotPrimes}. Similarly, approximations to Thomae's `ruler function', an exemplary pathological function of real analysis \cite{Burn}, have been measured in visible light optics \cite{ThomaeExp}. Factorization of a composite number using Gauss sums is also possible using Young's N-slit diffraction \cite{YoungNSlit} and has been demonstrated in Michelson interferometer experiments \cite{MichelsonGauss}. For an initial wave with Fourier transform proportional to the Riemann zeta function on the critical line, Berry has constructed far-field radiation patterns with side-lobes separated by the Riemann zeros \cite{BerryRiemannI, BerryRiemannII}.

In this Letter we are interested in whether simple wave superposition can give rise to the prime number sequence, in the absence of dynamical chaos or dedicated factorization checks.  We show that basic diffraction can sieve all multiples of composite numbers and thereby holographically encode the sequence of primes into a propagating wave-field. As such, these symmetric superpositions provide insights behind the orchestrated irregularity of the prime number sequence. Since the naturally diffracted fields are not defined by an algorithm, there are no sequential parameter adjustments and we do not exploit Gauss sums for factorizing specific composite numbers. Our construction is not based upon the Riemann zeta function or the Riemann hypothesis. The simplicity of this diffraction approach is exemplified by readily encoding other important sequences into propagating wave-fields, such as Gaussian primes, square-free integers, twin-primes and so on, as explained hereafter.

Wilson's theorem, that a prime $p$ divides $(p-1)!+1$ \cite{Havil}, could be employed to design a field with amplitude $\cos[\pi(x-1)!/x+\pi/x]$, the integer floor of which provides an indicator function for primes over integer $x$ positions. Whilst oscillatory, the construction is nonetheless contrived. A step further could be to consider the product $\prod_{n=2} \mathrm{sinc}(\pi x/n-\pi)$, which identifies all composite integer (non-prime) $x$ as a zero, since each sinc factor eliminates multiples of all but one integer of interest. Though based upon single-slit diffraction, it is difficult to envisage an experimental implementation of this series. We shall instead consider much simpler wave superpositions, which could be realized in an experiment -- essentially the opposite of the sinc idea, whereby composites are discarded by removing zeros in the wave field. 

\begin{figure}[htbp]
\includegraphics[width=1.0\linewidth]{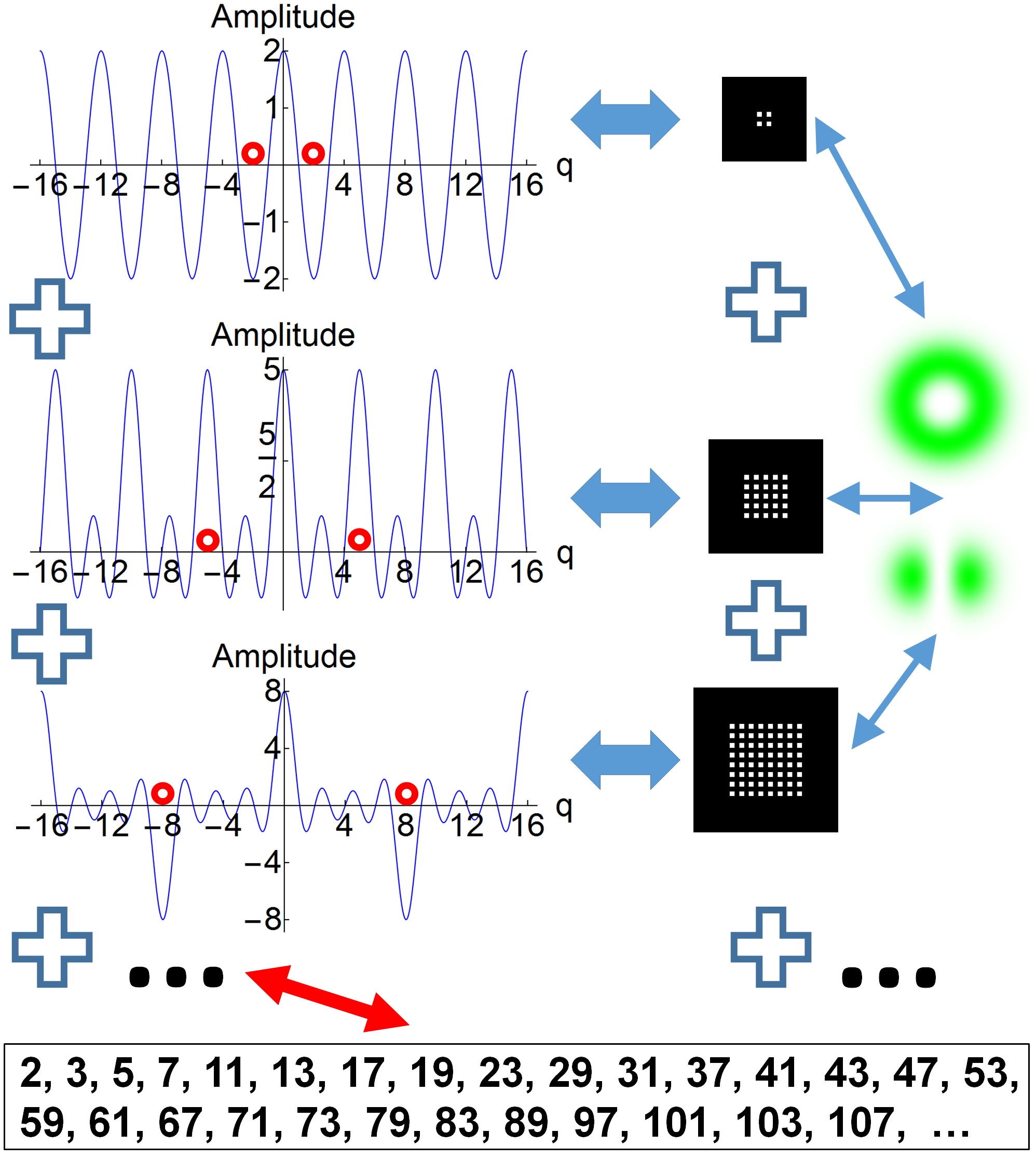}
\caption{Symmetric sets of waves encode the prime-number sieve of Eratosthenes. Diffraction of identical wave sources arranged in simple $N \times N$ grids creates interference patterns with zeros at all integer momenta $q$, except at multiples of $N$, where the amplitude equals $\pm N$. Hermite-Gauss mode sources (shown in green) can be used to place additional zeros at one or both of the red circles. Superposition then encodes primes as zeros in the total wave amplitude for integer $q$.}
\label{fig:Figure1}
\end{figure}

Suppose a superposed set $\Psi_N(\bm{r})$ of $N\times N$ identical scalar wave sources is located in the $x-y$ plane, with each source defined by wavefunction $\psi(\bm{r})$ at position $\bm{r}_j = (x_j,y_j)$; here the optic axis is along the $z$ direction. Far-field diffraction of $\Psi_N(\bm{r})$ can be written as the Fourier transform of $\psi(\bm{r})$ convolved with a set of Dirac deltas $\delta(\bm{r}-\bm{r}_j)$, i.e., $\widehat{\Psi}_N(\bm{q})=F[\psi(\bm{r})\ast\Sigma_j\delta(\bm{r}-\bm{r}_j)]$, where the spatial frequency or momentum is denoted by $\bm{q} = (q_x,q_y)$. For an $N\times N$ diffraction grating with $x_j = (j-(N+1)/2)/N$ (for $j = 1,2,3...N$), in dimensionless units, and likewise for the $y$ positions, the far-field wave is proportional to,
\begin{equation} 
\widehat{\Psi}_N(\bm{q})=\widehat{\psi}(\bm{q})\frac{\sin(\pi q_x)\sin(\pi q_y)}{\sin(\pi q_x/N)\sin(\pi q_y/N)},
\label{Eq1}
\end{equation}
where $\widehat{\psi}(\bm{q})=F[\psi(\bm{r})]$ is the Fourier transform of $\psi(\bm{r})$. For ideal pinholes, $\widehat{\psi}(\bm{q})$ tends to a constant and $\widehat{\Psi}_N(q_x,0)$ or $\widehat{\Psi}_N(0,q_y)$ then has form matching one of the graphs in Fig.~\ref{fig:Figure1} for a given $N$.

Inspection of Fig.~\ref{fig:Figure1} shows that $\widehat{\Psi}_N(\bm{q}) = \pm N$ when $q_x$ or $q_y$ is a multiple of $N$; the wave amplitude is otherwise zero at all other integer $q$ values. This is consistent with l'H\^{o}pital's rule, which yields the limiting value $N (-1)^{q_x(1+1/N)}$, for integer $q_x$ divisible by $N$ when $q_y$ is zero and vice versa. Along either $\bm{q}$-axis, Eq.~\ref{Eq1} can hence be viewed as a \textit{ruler for discrete momenta}, with non-zero markings at integer multiples of $N$. Superposing many different $\widehat{\Psi}_N(\bm{q})$ over a range of $N$-values creates Moir\'e patterns at integer momenta, since the amplitude is only non-zero at these points for periodic multiples of each $N$. This is similar to Eratosthenes' scheme for eliminating composite numbers: if the natural numbers are associated with discrete momenta, non-zero amplitudes along either of the $\bm{q}$-axes indicate that a trial integer momentum of interest is composite. Closer correspondence is assured if additional zeros can be inscribed in each $\widehat{\Psi}_N(\bm{q})$ at $q_x = N$ or $q_y = N$, for then a candidate prime $q$ location, on the particular $\bm{q}$-axis, remains zero if the momentum $q$ is not divisible by any of the $N$ in the sum over all $\widehat{\Psi}_N(\bm{q})$. Such additional isolated zeros can be incorporated if the $\psi(\bm{r})$ sources are simple modes of the paraxial Helmholtz equation.

Each temporal frequency $\omega$ component $\Psi_\omega (\bm{r},z) = \exp(2\pi i k z)\psi(\bm{r},z)$ of a paraxial scalar wave satisfies the paraxial Helmholtz equation $\{\partial_x^2+\partial_y^2+2ik\partial_z\}\psi(\bm{r},z)=0$ \cite{PagBook}, where $k$ is the wave number. Among plane waves and other forms, exact solutions of this equation are given by the Hermite-Gauss modes, which are also eigenfunctions of the quantum harmonic oscillator. In the $x-y$ plane, up to a complex constant, the (0,0), (1,0) and (0,1) order modes can be written as $\psi_{00}(\bm{r}) = \exp(-r^2/\sigma^2)$, $\psi_{10}(\bm{r}) = x\psi_{00}(\bm{r})$ and $\psi_{01}(\bm{r}) = y\psi_{00}(\bm{r})$, respectively, where $r$ = $|\bm{r}|$ and $\sigma$ is the beam waist at the plane $z = 0$. A superposition of (0,2) and (2,0) modes gives the quadratic form $\psi_{Q}(\bm{r})= r^2\psi_{00}(\bm{r})$. Any of these waves in the limit ${\sigma \to \infty}$ would suffice for this discussion, which correspond to `polynomial waves' \cite{PolyParaxDennis, PolyPaganin}.

For bounded $f(x)$ in the implicit Fourier convention of Eq.~\ref{Eq1}, the relation $F[\partial_x f(x)] = 2\pi i q_x F[f(x)]$ holds. Similarly, $F[x f(x)] = i/(2\pi) \partial_{q_x} F[f(x)]$. As such, $F[\psi_{10}(\bm{r})]$ contains a zero at the $q_x$ origin, which can be shifted along $q_x$ by $N$ units of momenta after applying a linear phase-ramp $\exp(2\pi i x N)$ to tilt the $N\times N$ wave array. Equivalent remarks hold if the identical wave sources are instead chosen to be $F[\psi_{01}(\bm{r})]$. Phase ramps for each $N\times N$ array could be difficult in an experiment but can be avoided by using the second-order mode sources, since $F[\psi_{Q}(\bm{r})]$ contains a ring of zeros in the far-field, due to $2^{nd}$ order gradients arising from the quadratic term $r^2$. By careful choice of $\sigma$, the ring radius can be set to $N$ units of momenta.

The classical Eratosthenes algorithm iteratively eliminates composites by crossing out multiples of primes identified at each step. Our wave-optical approach automatically sieves all integer multiples and is hence less efficient. Nonetheless Eq.~\ref{Eq2} represents a close analogy, 
\begin{equation} 
\widehat{\Psi}_M(\bm{q})=\sum\limits_{N=2}^{M}\widehat{\psi}_N(\bm{q})\frac{\sin(\pi q_x)\sin(\pi q_y)}{\sin(\pi q_x/N)\sin(\pi q_y/N)},
\label{Eq2}
\end{equation}
where $\widehat{\psi}_N(\bm{q})$ = $(-1)^N\widehat{\psi}(q_x-N,q_y)$, for a phase shifted array when each source is of type $\psi_{10}(\bm{r})$, or $\widehat{\psi}_N(\bm{q})$ = $\widehat{\psi}_{Q}(\bm{q})$ for the second-order mode sources of width $\sigma_N$ without a phase ramp. Using these superpositions, zeros in the far-field pattern at integer momenta on the $q_x, q_y$ axes indicate primes for integers $|\bm{q}|\le M^2$, with non-zero amplitudes specifying composite numbers. Sifting occurs indefinitely beyond $|\bm{q}|=M^2$ but some composite momenta will then also correspond to zero amplitudes. For a given signal-to-noise ratio, this sifting is physically limited by finite energy, giving rise to diminishing wave amplitudes at large  $|\bm{q}|$.

\begin{figure}[htbp]
\includegraphics[width=\linewidth]{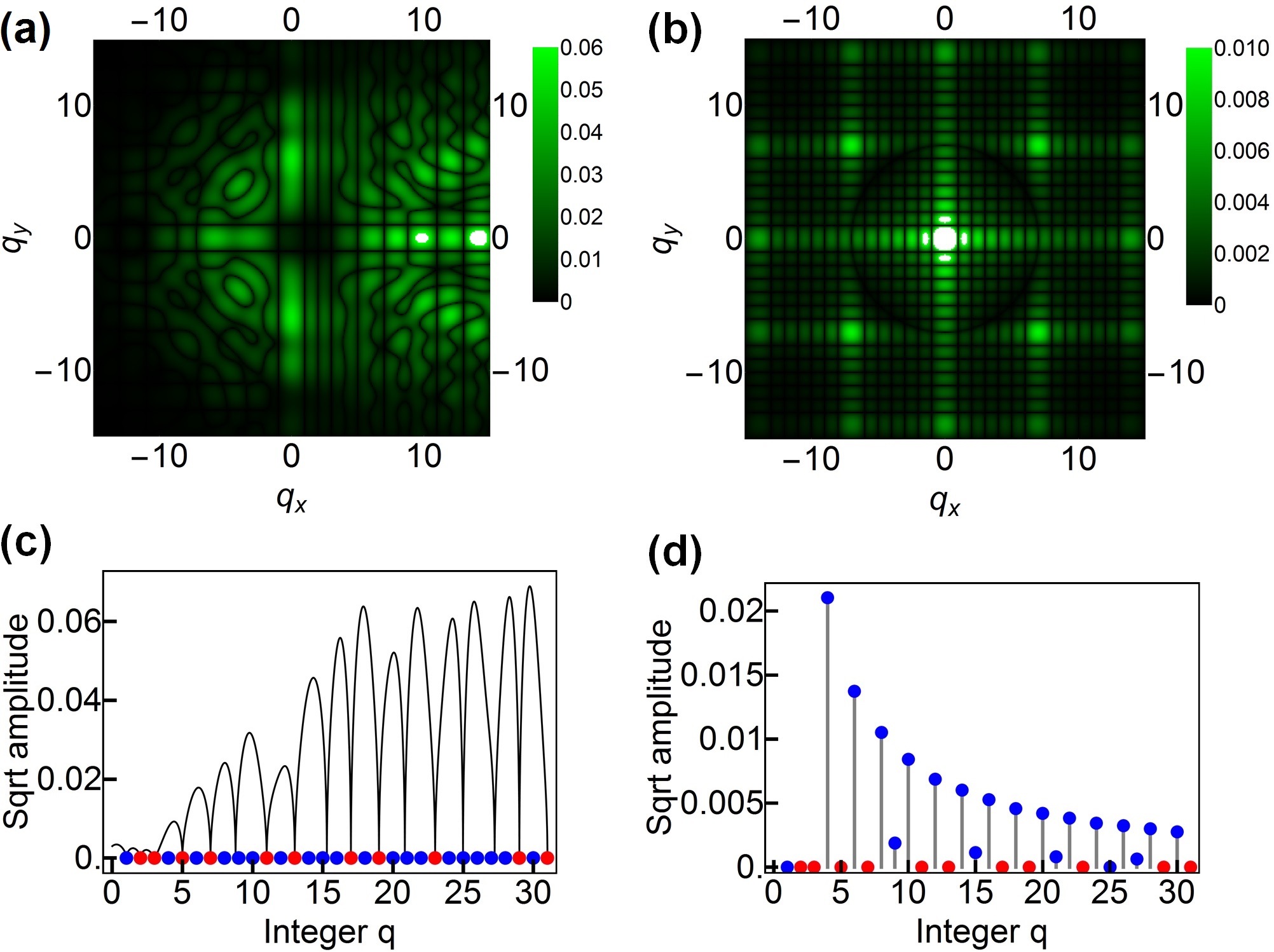}
\caption{The prime sieve of Eratosthenes realized with superposed Hermite-Gauss modes. (a) Square root of the far-field amplitude for phase-shifted first-order modes, with prime number zeros indicated as red dots for the corresponding $|\bm{q}| = q$ trace from the origin along $q_y = 0$ in (c). (b) Ring of zeros for a quadratic second-order mode with dimensionless radius $q=7$, superpositions of which yield the prime sieve in (d), for integer $q$ along either $\bm{q}$ axis.}
\label{fig:Figure2}
\end{figure}

Displaying the square root of the amplitude, Fig.~\ref{fig:Figure2} shows some examples of Eq.~\ref{Eq2}, with $\psi_{10}(\bm{r})$ and $\sigma = 0.05$, $M = 31$, producing Fig.~\ref{fig:Figure2}(a) and the corresponding trace along $q_y=0$ in Fig.~\ref{fig:Figure2}(c). The ring of size $|\bm{q}|=7$ in Fig.~\ref{fig:Figure2}(b) shows an example of $F[\psi_{Q}(\bm{r})]$ for $\sigma = 1/(N\pi)$ where $N=7$ and Fig.~\ref{fig:Figure2}(d) shows a horizontal trace over discrete $q_x$ computed from Eq.~\ref{Eq2} using $\psi_{Q}(\bm{r})$ sources with the sum maximum $M=31$. Note the almost zero values for even integers, non-integers, and $q_x$ = 25 in Fig.~\ref{fig:Figure2}(c), which is masquerading as a possible prime. The amplitude at $q_x$ = 25 evaluates to the tiny value of $24/(25 e^{25} \pi^3)$ but is strictly zero for prime $q_x$. These observations reveal important physical limitations for realizing wave-optical prime sieves. Similar remarks hold for the discrete plot in Fig.~\ref{fig:Figure2}(c), for an un-shifted $\widehat{\psi}_Q(\bm{q})$ based source spectrum.

Simpler wave-optical prime sieves are possible if mimicry of Eratosthenes' algorithm is jettisoned. Given that the trigonometric ratios in Eq.~\ref{Eq1} provide the essential divisibility tests for $q_x$ or $q_y$, it is instructive to devise more basic complex exponential sums over ordered rational frequencies. To this end, consider
\begin{equation} 
\widehat{S}(\alpha)\equiv\sum\limits_{N=1}^{M}\sum\limits_{j=1}^{N}e^{2\pi i \alpha j/N} = \sum\limits_{N=1}^{M}(-1)^{\alpha(1+N)/N}\frac{\sin(\pi\alpha)}{\sin(\pi\alpha/N)},
\label{Eq3}
\end{equation}
\noindent where $\alpha$ and $j/N$ could correspond to respective momentum and position or vice versa. Note that the sign alternation for $\alpha$ divisible by $N$ cancels that of the sine term, hence Eq.~\ref{Eq3} is a simpler one-dimensional version of Eq.~\ref{Eq1}, which can be furnished with a given source term $\psi(\alpha)$ through convolution, if relevant. From a diffraction physics perspective, the inner sum in Eq.~\ref{Eq3} can be viewed as a phase singularity \cite{Visser} for integer $\alpha$ not divisible by $N$, since the phasor sum inscribes a circle as a regular polygon in the Argand plane, winding $\alpha$ times about this polygon. When $N$ divides $\alpha$, the inner sum instead represents a plane wave of amplitude $N$, since all phasors add along a line in the Argand plane. As with all sums here, only the real part of the superposition is of physical significance \cite{BornWolf}.

The number-theoretic properties of Eq.~\ref{Eq3} are interesting. The inner sum evaluates to $N$ for $\alpha$ divisible by $N$, and is zero for all other integer $\alpha$. $\widehat{S}(\alpha)$ is therefore the divisor function $\sigma_{1}(\alpha)$ from number theory, describing the sum of all integers that divide integer $\alpha$. Equation~\ref{Eq3} thus represents another type of sieve, for which candidate prime $\alpha$ are identified as the fixed points $\widehat{S}(\alpha) = \alpha$. An interpretation in terms of Thomae's ruler function \cite{Burn} can also be made for ideal pin-holes, as explained in the Supplemental Material. In short, the distribution of the primes is given by the fixed points in the spectrum of Thomae's function for integer $\alpha$.

\begin{figure}[htbp]
\includegraphics[width=0.9\linewidth]{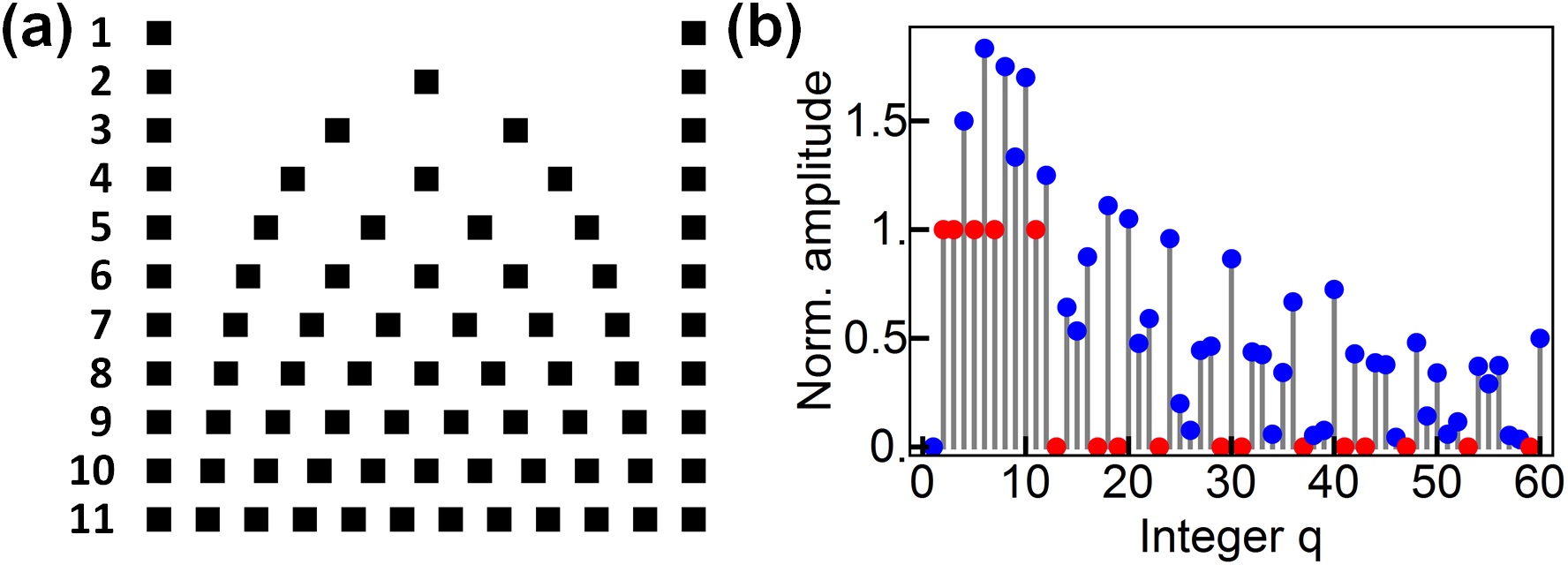}
\caption{(a) Primes encoded in a single symmetric aperture. (b) Normalized far-field diffraction from pinholes at the square locations in (a) identifies primes as unit amplitudes along the $q_x$ axis for momentum $|\bm{q}| = q \le 11$, beyond which zeros uniquely identify primes until $q>11^2$.}
\label{fig:Figure3}
\end{figure}

Superpositions such as $\widehat{S}(\alpha)$ can be realized in simplified diffraction experiments. For example, identifying $\alpha'$ with position $x$, the set of wave sources sampled by the 2D Dirac distribution $\sum\sum\delta(x-j/N)\delta(y-N)$ produces a far-field diffraction pattern matching Eq.~\ref{Eq3}, along the $q_{x}$ axis for $q_{y} = 0$. Extra $j = 0$ terms were included in Eq.~\ref{Eq3} for aesthetic purposes to plot this distribution as the set of black squares in Fig.~\ref{fig:Figure3}(a), where the numbers indicate the value of $N$ in the sum ($y$ points down the page), which ranges up to maximum $M = 11$ (cf. the set of natural line angles in the discrete Hough transform \cite{Imants}). Figure~\ref{fig:Figure3}(a) alone reveals the asymptote towards Thomae's function in the effective superposition, since there are roughly $1/2$ as many vertically aligned sources in the middle than the outermost columns, $1/3$ as many sources at either one $3^{rd}$ or two $3^{rd}$s of horizontal distance $x$, and so on. The Fourier transform of this distribution of ideal pinholes was computed and $M + 1$ was subtracted from the wave amplitude to account for the additional sources arising from the $j = 0$ terms. Further division by $q_{x}$ created the plot of normalized wave amplitude over discrete momenta $q\equiv q_{x}$ (at $q_{y} = 0$) shown in Fig.~\ref{fig:Figure3}(b), where the red dots are primes. As expected, the unity values arise from fixed points in the normalized divisor function $\sigma_{1}(q)/q$ up until $q = 11$. For this chosen value of $M$, primes are also uniquely identified by zeros in the wave field at integer $q$ up until $q = M^2 = 121$, since there are no terms in the effective $\widehat{S}(q)$ sum to contribute non-zero amplitude at discrete momenta $q$. Beyond this $q$, composite momenta can also give rise to zeros and the sieve no longer faithfully identifies candidate prime $q$. 

Small variations of Eq.~\ref{Eq3}, such as the inclusion of source types other than ideal pinholes, index changes, or reinterpretations of $\alpha$, can be used to adapt such wave sieves for other number theoretic wave fields of interest. For example, any integer can be represented as a unique product of squared and square-free integers, where a square-free integer contains no squares in its prime factor decomposition. By considering only rows $N = 4, 9...$ in Fig.~\ref{fig:Figure3}(a), the partial set of source locations then sifts integer momenta by the squares of all primes, resulting in zeros in the far-field diffraction pattern at integer momenta corresponding to all square-free numbers \cite{TwinPrimes}.

An acoustic example is a vibrating pipe open at both ends. With $\alpha$ identified as position $x$ along the pipe, the real part of Eq.~\ref{Eq3} describes the longitudinal wave displacement as a set of modes $\cos[\pi x q_{jNM}/(2L_{M})]$ for a pipe of length $L_{M} = \mathrm {LCM}(1,2,3...M)$ and particular harmonics $q_{jNM} = 4 j L_{M}/N$, where LCM is the least common multiple. At time $t=0$ for $x \le M$, fixed points of the longitudinal wave displacement identify prime $x$. For $M<x \le M^2$, zero wave displacement indicates prime $x$. Dynamics can be included with, say, a linear dispersion relationship. The displacement $x$ at a pipe end would be prime at prime instances of time, in units scaled by the dispersion relation.

Other connections between basic diffraction physics and number theory are possible with further variants of Eq.~\ref{Eq2} or Eq.~\ref{Eq3}. Two final number theoretical wave fields are worth discussing - a superposition containing the set of Gaussian primes and another that sieves twin primes.

\begin{figure}[htbp]
\includegraphics[width=\linewidth]{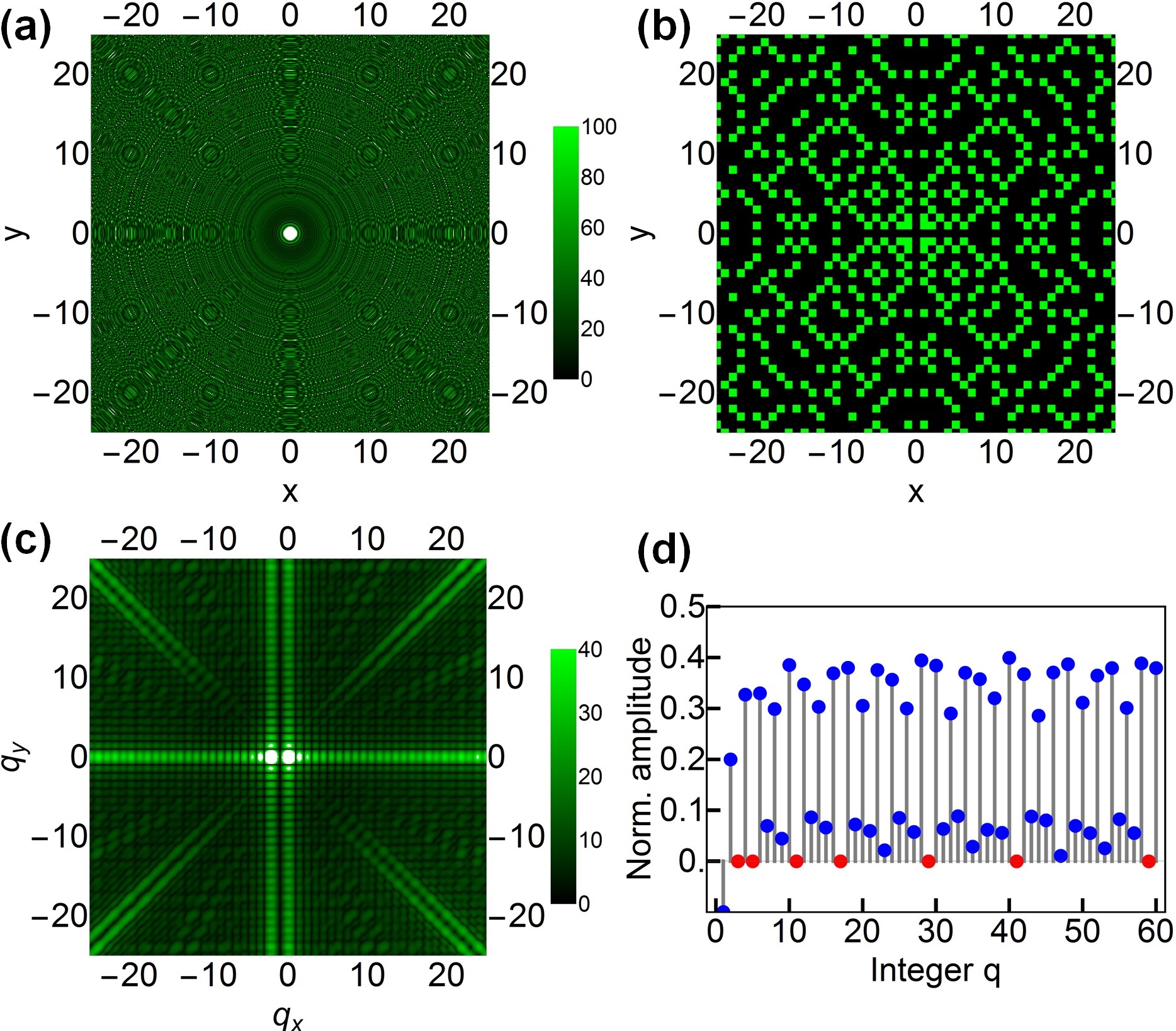}
\caption{Gaussian and twin primes (a) The magnitude of Eq.~\ref{Eq3} for $\alpha = r^2$. (b) Fixed points in (a) uniquely identify primes of form $x^2+y^2$. (c) Twin primes sieved by superposing diffraction patterns separated by 2 units of momenta along the $q_x$ axis using Eq.~\ref{Eq2}, plotted as the square root of the intensity. (d) Normalized plot over discrete $q$ along the $q_x$ axis, where the $1^{st}$ of each twin prime is uniquely zero, shown in red.}
\label{fig:Figure4}
\end{figure}

For integers $a$ and $b$, the Gaussian integers $a + ib$ are complex numbers which can be uniquely factorized by other Gaussian integers known as `Gaussian primes' that have norm $a^2 + b^2$ equal to a prime number \cite{PrimeBook}. When $\alpha$ in Eq.~\ref{Eq3} is interpreted as $r^2$ = $x^2+y^2$, this superposition represents a sum of paraxial spherical waves on the optic axis $(z = 0)$, which automatically sifts Gaussian primes for integer $x$ and $y$. The Gaussian sieve continuum in Fig.~\ref{fig:Figure4}(a) was computed from Eq.~\ref{Eq3} up to $M = 23$, ploting the square root of the intensity. The field is most intense at the origin, since all sources lie on the optic axis. The bright squares in Fig.~\ref{fig:Figure4}(b) are the same data, where each square shows the fixed points of the wave at integer $(x,y)$ positions, corresponding precisely to all Gaussian primes. In experiment, the various $j/N$ phase curvature factors could arise from different source locations on the optic axis by extending Eq.~\ref{Eq3}, as shown in the Supplemental Material.

Any of our sieves can be applied concurrently to remove multiple distributions of integers, using the superposition principle. For example, twin primes were sieved by adding Eq.~\ref{Eq2} to an identical wave shifted along $q_x$ by two integer units of momenta. The wave $\widehat{\psi}_N(\bm{q})$ was set to unity for simplicity to yield the square root of intensity shown in Fig.~\ref{fig:Figure4}(c), with $M=62$ in Eq.~\ref{Eq2}. The corresponding graph over discrete $q_x$ in Fig.~\ref{fig:Figure4}(d) was computed after normalizing the wave magnitude by $q_x^2+(q_x+2)^2$ and subtracting one, such that twin primes appear as unique integer zeros. Generalizations to sieve other prime gaps, tuples etc. are possible. While finite energy constraints are fundamental, technical issues such as non-paraxial diffraction can be overcome as Eq.~\ref{Eq3} can be viewed as a simple sum over plane waves.  

In conclusion, a wide variety of prime number sieves has been demonstrated using simple wave superposition. Examples were chosen to easily locate prime numbers in frequency, space and time, placing constraints on the architecture of the source distributions. Given natural phenomena such as beats and modes, wave fields are littered with integers in general, so perhaps the set of prime numbers implicitly resides within more general wave-field superpositions.

T.C.P. acknowledges useful discussions with A.C.Y. Liu.

%
\newpage
\section{\label{sec:level2}Suppl. Mat.: The diffraction spectrum of Thomae's function } Consider the function $
T_{M}(\alpha') \equiv F^{-1}[\widehat{S}(\alpha)],$
with the inclusion of wave sources $\psi(\alpha')$ normalized by $M$, where $\alpha'$ is Fourier-conjugate to $\alpha$. The Fourier shift theorem gives 
\begin{equation}
T_{M}(\alpha') = M^{-1} \psi(\alpha')\ast\sum\sum\delta(\alpha'-j/N), \nonumber
\end{equation} 
where the sums range as in Eq.~3 of the main text and $\ast$ denotes convolution. Each $M^{th}$ row of these delta functions, evenly spaced on the unit interval, can be written as $\Pi(\alpha')\mathrm{III}_{N}(\alpha')$, where $\Pi(\alpha')$ is the unit step function and $\mathrm{III}_{N}(\alpha')$ is a Dirac comb with teeth separated by $1/N$ units on the $\alpha'$ axis. The inverse transform is then  $\psi(\alpha')\ast\{M^{-1}\Pi(\alpha')\sum \mathrm{III}_{N}(\alpha')\}$, where the sum runs from $N$ to $M$. Consider now a rational point $\alpha' = n/m$ with co-prime $(n,m)$, such that $M$ is divisible by $m$ and choose ideal pin-holes $\psi(\alpha')\to\delta(\alpha')$. Observe that 
\begin{equation}
\mathrm{III}_{N}(n/m) = \mathrm{III}_{N}(1/m), \nonumber
\end{equation} 
since scaling $1/m$ by $n$ simply selects another tooth of the comb or else there are no teeth at either location and both sides are then zero. If $M$ is divisible by $m$, the $\mathrm{III}_{N}(n/m)$ term gives unity for $m = N = M$, as the fraction $1/m$ uniquely identifies just one tooth out of all combs in the double sum $F^{-1}[\widehat{S}(\alpha)]$. This is true for all $m$ when $M$ tends to infinity, hence 
\begin{equation}
\lim_{M\to\infty} T_{M}(n/m) = 1/m. \nonumber
\end{equation} 
The definition of Thomae's function is that $T(n/m) = 1/m$ for co-prime $(n,m)$ and zero for irrational arguments, which implies that 
\begin{equation}
\lim_{M\to\infty} T_{M}(\alpha') = T(\alpha'). \nonumber
\end{equation}

\section{\label{sec:level1} Suppl. Mat.: Gaussian prime sieve from a simple longitudinal grating}

Consider near-field diffraction of a unit-amplitude plane wave through an ideal pin-hole on the optic axis, with the phase advanced to zero at the observation plane. The diffracted field is proportional to $\exp[2 \pi i r^2/(2\lambda z)]/(\lambda z)$, after Fresnel propagation by the distance $z$. The scattered wave from a complementary aperture, such as an opaque infinitesimal particle, is $1-\exp[2 \pi i r^2/(2\lambda z)]/(\lambda z)$, by Babinet's principle [20]. Consider now a distribution of such sources at non-overlapping inverse distances 
\begin{equation}
1/(2\lambda z) = j/N+N \nonumber
\end{equation}
on the optic axis. By analogy with Fig.~3(a), this diffraction is from a single longitudinal grating. It can be shown that \begin{equation}
\sum2(j/N+N) \cos[2 \pi r^2(j/N+N)] - 1 = N(2N+1) \nonumber
\end{equation}
 when $N$ divides $r^2$ and zero for all other dimensionless $r^2$ integers. For a successive sum over all $N$, kinematical diffraction of a plane wave from this set of point-particles on the optic axis creates a near-field prime sieve superposition with wave amplitude proportional to $\sigma_{1}(r^2)+2\sigma_{2}(r^2)$, where $\sigma_{2}(\alpha)$ is the second-order divisor function (the sum of squares of all factors of $\alpha$). Normalization of the amplitude by $r^2(2r^2+1)$ gives unity for Gaussian prime $r^2$ and larger values for all other integer $r^2$. Explicit scattering cross sections for identical scatterers can be included by convolution of a suitable function in the transverse $(x,y)$ coordinates, or $r$ for radially symmetric particles.


\begin{thebibliography}{99}

\bibitem{Susan} S. H. Marshall and D. R. Smith, Math. Mag. \textbf{86}, 189 (2013)

\bibitem{RiemannReview}
D. Schumayer and D. A. W. Hutchinson, Rev. Mod. Phys. \textbf{83}, 307 (2011).

\bibitem{Havil} J. Havil and F. Dyson, \textit{Gamma: Exploring Euler's Constant}, (Princeton University Press, Princeton and Oxford, 2003).

\bibitem{PrimeBook} B. Mazur and W. Stein, \textit{Prime Numbers and the Riemann Hypothesis}, (Cambridge University Press, Cambridge, 2016).

\bibitem{BilliardsBrunimovich} L. A. Bunimovich and C. P. Dettmann, Phys. Rev. Lett. \textbf{94}, 100201 (2005).

\bibitem{BerryKeating} M. V. Berry and J. P. Keating, SIAM Rev. \textbf{41}, 236 (1999).

\bibitem{Bender} C. M. Bender, D. C. Brody and M. P. M\"{u}ller, Phys. Rev. Lett.  \textbf{118}, 130201 (2017).

\bibitem{CantorSet} S. Sears, M. Soljacic, M. Segev, D. Krylov and K. Bergman, Phys. Rev. Lett. \textbf{84}, 1902 (2000).

\bibitem{BerryKlein} M. V. Berry and S. Klein, J. Mod. Opt. \textbf{43}, 2139 (1996).

\bibitem{Talbot} C. R. Fern\'{a}ndez-Pousa, J. Opt. Soc. Am. A \textbf{34}, 732 (2017).

\bibitem{TalbotPrimes} K. Pelka, J. Graf, T. Mehringer and J. von Zanthier, Opt. Express \textbf{26}, 15009 (2018).

\bibitem{Burn} R. P. Burn, \textit{Numbers and Functions: Steps Into Analysis, 2nd Ed.}, (Cambridge University Press, Cambridge, 2000).

\bibitem{ThomaeExp} V. Saveljev and S.-K. Kim, Opt. Express \textbf{23}, 25597 (2015).

\bibitem{YoungNSlit} J. F. Clauser and J. P. Dowling, Phys. Rev. A, \textbf{53}, 4587 (1996).

\bibitem{MichelsonGauss} V. Tamma, H. Zhang, X.  He, A. Garuccio, W. P. Schleich and Y. Shih, Phys. Rev. A \textbf{83}, 020304(R) (2011). 

\bibitem{BerryRiemannI} M. V. Berry J. Phys. A: Math. Theor. \textbf{45}, 302001 (2012)

\bibitem{BerryRiemannII} M. V. Berry J. Phys. A: Math. Theor. \textbf{48}, 385203 (2015)

\bibitem{PagBook} D. M. Paganin, \textit{Coherent X-Ray Optics}, (Oxford University Press, Oxford, 2006).

\bibitem{PolyParaxDennis} M. R. Dennis, J. R. G\"{o}tte, R. P. King, M. A. Morgan and M. A. Alonso, Opt. Lett. \textbf{36}, 4452 (2011).

\bibitem{PolyPaganin} D. M. Paganin, M. A. Beltran and T.  C. Petersen, Opt. Lett. \textbf{43}, 975 (2018).

\bibitem{Visser} G. Gbur, T. D. Visser and E. Wolf, Phys. Rev. Lett. \textbf{88}, 013901 (2001). 

\bibitem{BornWolf} M. Born and E. Wolf, \textit{Principles of Optics, seventh (expanded) edition}, (Cambridge Univ. Press, Cambridge, 1999).

\bibitem{Imants} I. D. Svalbe, IEEE Trans. Pattern Anal. Mach. Intell. \textbf{11}, 941 (1989).

\bibitem{TwinPrimes} H. Halberstam and H.-E. Richert, \textit{Sieve Methods}, (Academic Press, London, 1974).

\end{thebibliography}
\end{document}